# Generalized all-optical complex exponential operator

Baiqiao Chen, Qi Jia,* Rui Feng, Fangkui Sun, Yongyin Cao, Jian Wang,* and Weiqiang Ding*

B. Chen, Q. Jia, R. Feng, F. Sun, Y. Cao, W. Ding

Institute of Advanced Photonics, School of Physics, Harbin Institute of Technology, Harbin 150001, China

E-mail: jiaqi1350@163.com; wqding@hit.edu.cn

*J. Wang*

School of Physics, Harbin Institute of Technology, Harbin 150001, China

E-mail: hitwj@hit.edu.cn

W. Ding

Collaborative Innovation Center of Extreme Optics, Shanxi University, Taiyuan 030006, China

E-mail: wqding@hit.edu.cn

Euler's formula, an extraordinary mathematical formula, establishes a vital link between complex-valued operations and trigonometric functions, finding widespread application in various fields. With the end of Moore's Law, electronic computing methods are encountering developmental bottlenecks. With its enviable potential, optical computing has successfully achieved high-speed operation of designed complex numbers. However, the challenge of processing and manipulating arbitrary complex numbers persists. This study introduces a generalized complex exponential operator (GCEO), utilizing a diffractive optical neural network (DONN) for the computation of the complex exponential through Euler's formula. Experiments validate a series of complex exponential calculations using the GCEO. The GCEO has demonstrated generalizability and can compute inputs of any precision within an appropriate error margin. The proposed operator highlights the immense potential of DONN in optical computation and is poised to significantly contribute to the development of computational methods for optoelectronic integration.

# 1. Introduction

Euler's formula is widely recognized as one of the most elegant expressions in mathematics. It plays a central role in complex analysis, establishing a profound link between trigonometric and exponential functions. The formula's extensive applications span various fields, including applied mathematics,[1,2] electromagnetism,[3–7] quantum mechanics,[8–13] fluid mechanics,[14,15] and signal analysis,[16–18] highlighting its critical role in both theoretical research and engineering applications. Obviously, Euler's formula is integral to numerous computational tasks.

In Euler's formula, the interplay between real and imaginary components of complex numbers is pivotal, especially in computations involving complex exponentials in practical scenarios. Traditional electronic devices such as Digital Signal Processors and Field-Programmable Gate Arrays are typically used for these calculations. However, these electronic devices exhibit limited parallelism when executing computations. Moreover, in the post-Moore's Law era, they encounter challenges including high power consumption and increased complexity.[19] In contrast, optical computing, known for its inherent parallelism, rapid processing speeds, low power consumption, and high throughput, emerges as a promising solution.[20–25] Employing optical computing for specific intricate tasks, particularly those involving complex numbers, is a strategic approach to address these computational challenges.[26–28] Optical computing, utilizing phase and amplitude encoded information, is capable of performing trigonometric function calculations.[29] However, this operation lacks generalizability, with a limited range of inputs it can process effectively. Moreover, while operating trigonometric calculations for a limited number of angles is feasible, the challenge escalates when dealing with multiple trigonometric functions and their integration with complex exponents through Euler's formula.

In 2018, Lin et al. introduced the Diffractive Optical Neural Network (DONN),[30] providing a new approach to optical computing systems. This architecture is notably flexible, showing rapid advancements in image processing,[31–37] and even the ability to handle a broader range of complex tasks.[38–48] These developments vividly illustrate the significant potential of DONN in tackling complex computational challenges. It is worth noting that optical neural networks are capable of executing complex-valued matrix-vector multiplications.[49] However, this capability is confined to scenarios where the inputs are explicitly defined in terms of their real and imaginary components. In contrast, in practical applications, complex numbers are more commonly encountered in their exponential form. Additionally, optical neural networks can implement complex linear transformations,[50,51] yet the outcomes in terms of complex number calculations remain suboptimal, as the outputs do not fully represent complex values. Another strategy involves using high-dimensional optical angular momentum for encoding complex vectors and conducting convolution operations,[52] but this method is limited to a finite set of complex number operations and lacks optical recognition of computational results.

To overcome these limitations, we introduce a generalized complex exponential operator (GCEO) based on Euler's formula and implemented using DONN. Recognizing the distinctions between computation outcomes and typical classification operations in neural networks, we have developed an innovative approach utilizing the intensity modulation capability of DONN. This method leverages optical intensity representation for numerical calculations, encoding the computation results' real and imaginary components with different signs in distinct regions of the output plane. Our experimental setup consists of a three-layer DONN, comprising a spatial light modulator (SLM) and a mirror, which processes input beams at various spatial angles. Theoretically, we have formulated the GCEO, which we have experimentally validated for its capability to compute exponentials using angles between -$\pi$ and $\pi$. We are confident that this generalized operator will substantially advance the exploitation of optical complex number computations.

## 2. Methods

The convergence of DONN with Euler's formula presents intriguing opportunities for advanced complex computations. Bridging these two domains demands innovative approaches, blending DONN's proficiency in manipulating coherent light with the mathematical intricacies of Euler's formula. Addressing these challenges is essential for harnessing the full capabilities of optical computing in complex mathematical operations.

Without loss of generality, Euler's formula can be expressed in the following form:

$$e^{i\varphi} = \cos\varphi + i\sin\varphi \qquad (1)$$

For the design of a Diffractive Optical Neural Network (DONN) that executes complex exponential calculations using Equation (1), it is necessary to load the arguments of the complex exponential functions at the input plane. On the output plane, the real and imaginary parts of the results from Euler's formula computations should be appropriately displayed and distinguished by their positive and negative values. In our approach, Gaussian spots and encoded apertures were selected to form the training set for inputs and target outputs.

For the input training set, the angle ($\theta$) between Gaussian spots and a horizontal axis originating from the center of the plane was considered (e.g. 130°), as indicated in **Figure 1**. The $\theta$ range extended from -$\pi$ to $\pi$, with a one-degree increment between each angle, cumulating in 360 angles. In the target output training set, the computed outcomes of Euler's formula were encoded onto the power of selected apertures. Using the method of encoding computational results numerically through light energy ensures that the outputs are continuous, guaranteeing that the system can compute appropriate results for inputs of any precision. The output plane encoding is demonstrated in Figure 1. The upper and lower apertures, indicated by blue dashed lines, represent the imaginary components of the computed values, whereas the left and right apertures, shown with red dashed lines, indicate the real components.

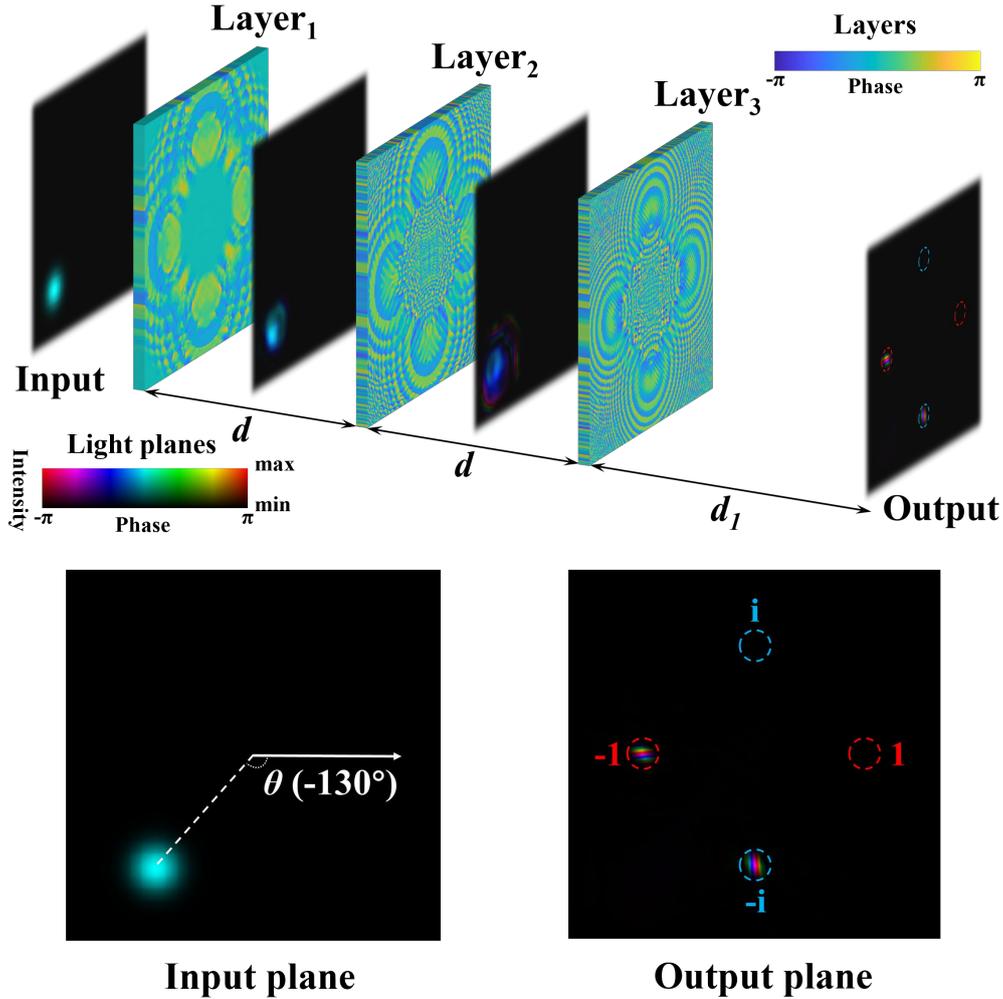

**Figure 1.** Schematic of Euler's formula computation via three-layered DONN. The setup incorporates three diffractive layers of $Layer_{1,2,3}$. The interlayer distance ($d$) is 7.95 cm, while the distance from the final layer to the output plane ($d_1$) measures 10.6 cm. To enhance visibility, enlarged views of the input and output planes are displayed below. On the input plane, a horizontal arrow emanating from the central point delineates the axis, with the angle ($\theta$) between this axis and the Gaussian beam illustrated, $\theta$ being 130° in this example. The output plane features the encoded apertures, demarcated by a white dotted line, representing the computation's results.

For the DONN, according to the angular spectrum theory, the propagation of an optical field can be described as,[53]

$$\begin{cases} E(x,y,z) = \hat{\mathcal{F}}^{-1}\left\{\hat{\mathcal{H}}\hat{\mathcal{F}}\left[E(x,y,0)\right]\right\} \\ \hat{\mathcal{H}} = \exp\left(iz\sqrt{k^2 - k_x^2 - k_y^2}\right) \end{cases} \qquad (2)$$

where $\hat{\mathcal{F}}$ is the operator for the two-dimensional Fourier transform, and correspondingly, $\hat{\mathcal{F}}^{-1}$ represents the inverse two-dimensional Fourier transform. $\hat{\mathcal{H}}$ is the transfer function. $k$ is the angular wavenumber of the light fields, and $k_x$ and $k_y$ are the projections along $x$ and $y$ directions, respectively.

In the case of finite sampling, the integral in Equation can be rearranged into the form of the matrix. for convenience, we define the diffractive matrix $D$ as

$$D = \mathcal{F}^{-1}\mathcal{H}\mathcal{F} \qquad (3)$$

where $\mathcal{F}$ is the two-dimensional discrete Fourier transform, and correspondingly, $\mathcal{F}^{-1}$ represents the inverse two-dimensional discrete Fourier transform. $\hat{\mathcal{H}}$ is the discrete transfer function.

For the building of the DONN, we use Equation (3) to simulate the forward propagation,

$$\begin{cases} X_{n+1}^{(\alpha)} = DT_n X_n^{(\alpha)} \\ T_n = \mathrm{diag}(\exp(i\varphi_n)) \\ X_{out}^{(\alpha)} = K X_{N_l+1}^{(\alpha)} \\ Y_{out}^{(\alpha)} = \mathrm{sum}\left(\left|X_{out}^{(\alpha)}\right|^2\right) \end{cases} \qquad (4)$$

where $X_n^{(\alpha)}$ is a column vector of the input light on the $n$-th diffractive layer for the $\alpha$-th sample. $\varphi_n$ is a column vector of the phase in the $n$-th diffractive layer. diag(·) represents the diagonalization operation. The total number of diffractive layers is set to be $N_l$. $X_{out}^{(\alpha)}$ is the actual output light in operator's apertures of the DONN for α-th sample. To implement the designed computational scheme, we incorporated apertures from Figure 1 into the training as mask versions. $K$ is a mask containing four apertures, used for statistical analysis of the output light field energy to produce computational results. The output plane in Figure 1 displays the mask $K$, where the red dashed lines represent the real part apertures, and the blue dashed lines represent the imaginary part apertures. $Y_{out}^{(\alpha)}$ is the actual output complex data of the DONN for α-th sample. sum(·) represents the addition operation.

Backward propagating and gradient descent algorithms are used to train the DONN. Without loss of generality, three neural layers of Layer$_{1,2,3}$ are set in for the DONN, comprising 300×300 optical neurons with each neuron measuring 8 × 8 μm, was employed for this operation. This neuron size corresponds to the pixel dimensions of the spatial light modulator (SLM) used in our experiments. The chosen wavelength ($\lambda$) was 532 nm, with inter-layer distances ($d$) set at 7.95 cm and the distance from the final layer to the output plane ($d_1$) measures 10.6 cm. Figure 1 illustrates the results post-training. The GCEO is capable of reflecting the results of Euler's formula complex exponential calculations through the energy information within four apertures on the output plane.

## 3. Results

Figure 1 demonstrates the successful computation of Euler's formula by the DONN post-training. Specifically, when the input beam angle $\theta$ is set at -130°, the output plane reveals that both the real and imaginary parts of the operation result are negative. The corresponding intensity of light accurately reflects these values.

To validate the practical performance of the GCEO, we established a corresponding experimental setup. **Figure 2**(a) depicts the schematic of this setup. The experimental process begins with a Gaussian laser beam of wavelength 532 nm, which is expanded and collimated through $Lens_1$, $Lens_2$, and $Iris_1$, followed by polarization via polarizer P. Subsequently, a forked phase hologram is applied on $SLM_1$ (Holoeye PLUTO-2-VIS-016) to produce a Gaussian spot with varying angles $\theta$.[54,55] A representative phase hologram pattern is showcased in Figure 2(b). After passing through pinhole filtering ($Iris_2$) and a 4-$f$ system ($Lens_3$ and $Lens_4$), the Gaussian beam enters the DONN, consisting of $SLM_2$ (Holoeye PLUTO-BB) and a plane mirror M. In this setup, $SLM_2$ is spatially multiplexed thrice using mirror M to construct a three-layered DONN. Each segment of $SLM_2$ is loaded with numerically solved phase patterns corresponding to the layers $L_1$ to $L_3$ seen in Figure 1. This configuration enables the use of a single SLM to simulate all three layers, offering considerable experimental convenience. The output light, representing the computational results, is captured by a CCD camera. Data is acquired by quantifying the light intensity within the encoded apertures.

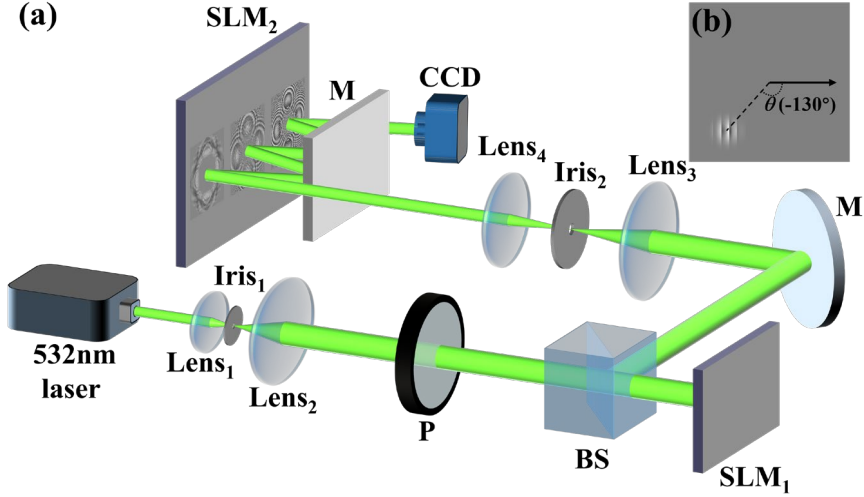

**Figure 2.** Experimental setup. (a) Schematic of the Experimental Setup for GCEO using a Three-Layered DONN. A Gaussian laser beam of wavelength 532 nm undergoes expansion and collimation via Lens$_1$ and Lens$_2$, followed by polarization through polarizer P. SLM$_1$ then transforms the Gaussian beam into a Gaussian spot with a predefined size and a series of plane angles $\theta$. Post the 4-$f$ system, the input encoded beams are processed by the DONN, comprising SLM$_2$ and mirror M. The output light, representing the computational results, is captured by a CCD camera. Data is acquired by quantifying the light intensity within the encoded apertures. (b) A phase hologram pattern used in SLM$_1$ for generating incident Gaussian beams with varying $\theta$. List of Abbreviations: Beam Splitter (BS); Spatial Light Modulator (SLM); Mirror (M); Polarizer (P).

To ensure that the input angle $\theta$ is accurately operated into its corresponding aperture to export precise output complex numbers, we introduce an error number $e_\alpha$ for measurement. The overall error $e$ is then given by $e = \frac{1}{N}\sum_\alpha e_\alpha$, where $e_\alpha$ is defined as follows,

$$e_\alpha = \left| X_{out}^{(\alpha)} - Y_{tar}^{(\alpha)} \right| \tag{5}$$

where $X_{out}^{(\alpha)}$ and $Y_{tar}^{(\alpha)}$ are the target output and the actual output of the DONN for $\alpha$-th sample, respectively.

**Figure 3** displays the experimental results of Euler's formula computation using this system. The experimental results are presented through the output light intensity within the aperture channels. The output intensity data from both simulation and experiments are normalized by dividing all collected data by their mean values.

The computational capability of the GCEO was evaluated by using 360 angles as the training set and expanding to 720 angles between -$\pi$ and $\pi$ as the test set. As shown in Figures 3(a) and 3(b), the simulation and experimental results display the output for various angles $\theta$ across different components. The black lines represent the actual target values, while the red and blue curves correspond to the real and imaginary parts of the GCEO's computed results, respectively. Notably, the intensity trends of the red real part apertures (-1 and 1) and the blue imaginary part apertures (-i and i) vary with the angles $\theta$ in a manner consistent with the sine and cosine functions. These observed trends align with the theoretical values predicted by Euler's formula.

It is evident that the GCEO's computational results from simulations are closer to the actual values, while the experimental conditions introduce some errors. Figures 3(c) and 3(d) illustrate the distribution of computational errors from both the simulation and the series of experiments. These figures demonstrate the error patterns and their distributions, providing insights into the accuracy and reliability of the GCEO under different scenarios.

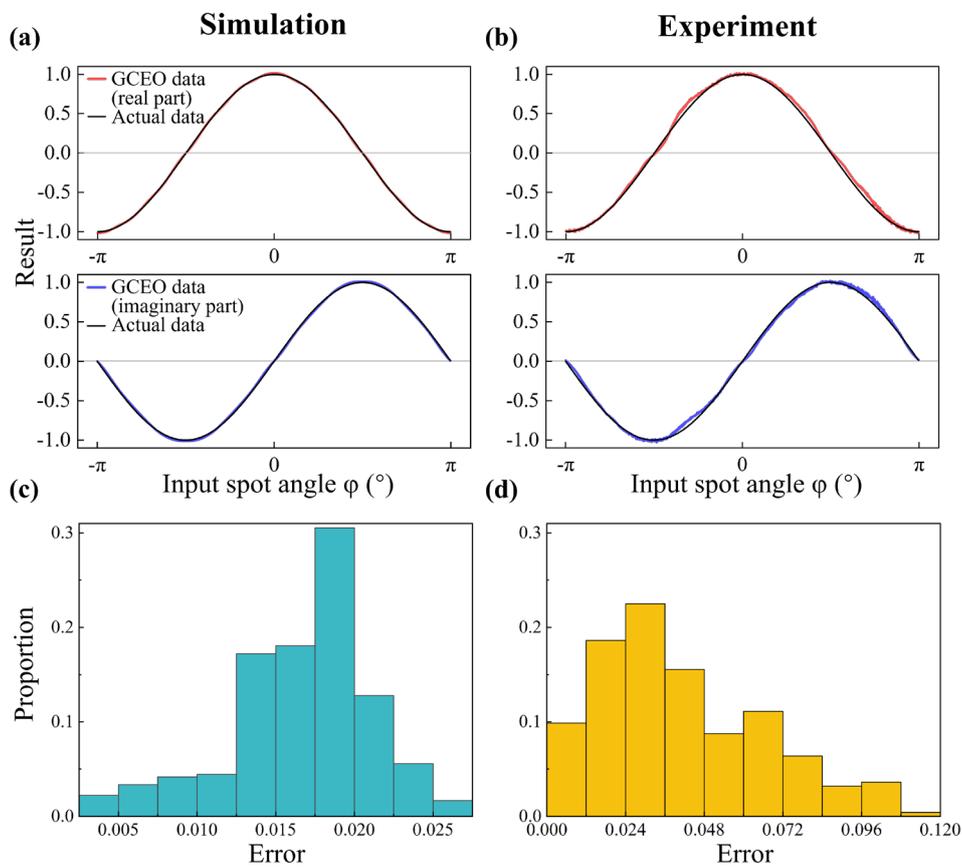

**Figure 3.** Test results of the GCEO. (a) and (b) show the computational outputs of the GCEO under different inputs angles $\theta$ in simulation and experimental conditions, respectively. These results are presented through the output light intensity within the aperture channels, normalized by the same exposure. The black curve represents the actual target values, while the red and blue curves correspond to the computed real and imaginary parts, respectively. (c) and (d) illustrate the error distribution histograms for simulation and experimental conditions, providing insights into the accuracy and reliability of the GCEO under different settings.

The average computational error in the simulation tests was 0.0167, as depicted in Figure 3(c). During experimental validation, the accuracy of SLM's modulation influenced optical losses in different regions, leading to deviations in total output energy for each input position compared to the uniformity observed in simulation tests. Consequently, the light intensity of the calculated results for each channel varied. The increase in optical losses led to a greater error in the experimental results. Figure 3(d) shows the error distribution of the experimental calculations, with an average error of 0.0412.

To enable the calculation system to output results directly, thus avoiding the need for further normalization, we quantified the optical output loss in each region of the input system. We then compensated for the input light intensity to mitigate the excess energy loss caused by experimental errors. With this approach, the output energy corresponding to different inputs became nearly identical, allowing for direct reading of light intensity in various apertures. The average error under these conditions was 0.0412. The error distribution for this scenario is presented in Figure 3(d). In the experiments, we only applied uniform exposure to the output to ensure the overall output magnitude was around 1. If each output were individually normalized, the experimental error would be smaller compared to the data obtained using only exposure. The average error obtained using this normalization method was 0.0387. Our approach does not require post-processing of the output data; instead, uniform exposure of the output light suffices to achieve reliable computational results.

## 4. Discussion

The computational error associated with the GCEO remains within a specific range, ensuring reliable computations. Notably, the operator exhibits generalizability and is theoretically capable of computing any complex Euler formula. In simulations, the average error was reduced to less than 0.01, while in experimental settings, it averaged around 0.04. This level of accuracy holds considerable importance for the practical application of the GCEO. It should be noted that the primary source of error stems from accumulated aberrations during the passage through the SLM. In principle, these errors could be mitigated by generating compensatory phases through various methods,[56–58] or using the Zernike function to generate the compensated phase.

The generalizability of the GCEO fundamentally derives from the precision in the division of light energy ratios. Once the division density reaches a certain threshold, GCEO can accurately output results within the error margin for angular inputs not included in the training set. To evaluate the variations in GCEO's generalizability, we trained the operator using the same training parameters but with different training sets, and the average error was statistically analyzed using test sets of various sizes, with the results illustrated in **Figure 4**(a). The red dotted lines indicate the outcomes when the number of test sets matched the number of training sets, assessments of generalizability should focus on data beyond the red dashed markers. It is observed that the computational error of the network increases when tested with a test set that has a higher angular density than the training set, specifically under training with 16 and 32 inputs. Conversely, when the training set comprises 64 or more inputs, the average error remains nearly constant, even when subjected to a test set larger than the training set.

In addition to observing the onset of generalization with a certain number of training sets, it is noteworthy that the error significantly decreases when the training set size increases from 16 to 32. After achieving noticeable generalization with 64 training sets, further increases in the training set size result in relatively smaller reductions in error. Our training sets increase exponentially by a factor of 2, effectively inserting a new input between each pair of input points from the previous set. The change in training effectiveness can be explained by the number of neurons utilized in the computation. For instance, with 16 or fewer training sets, the angle range covered by the inputs is very limited, leaving large regions of the plane unoccupied. Consequently, the neurons corresponding to these regions in the neural layer remain underfitted. As the training set size increases, the input points more comprehensively cover the plane, enabling the previously underfitted neurons to achieve fitting.Moreover, the complete coverage of the input plane and the corresponding fitting of the neural layers mean that even minor changes in the input can be effectively responded to by the already fitted diffractive neural network. Thus, further increases in training set size yield diminishing improvements in error reduction once a certain level of coverage and fitting is achieved.

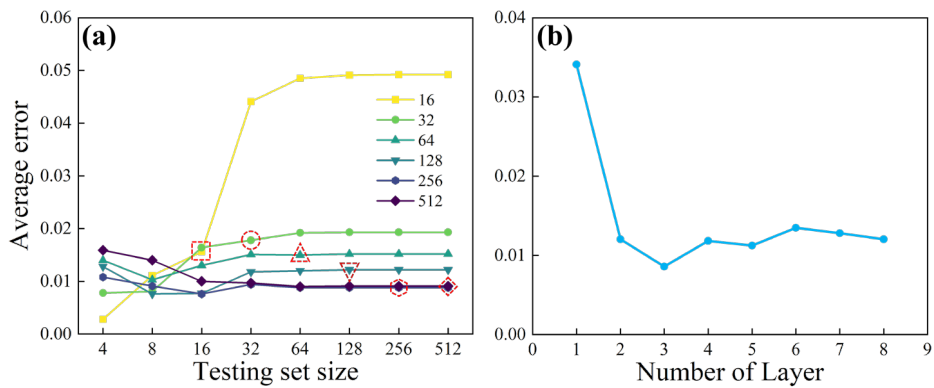

**Figure 4.** The impact of different DONN Parameters on GCEO computational error. (a) generalizability test results illustrate the average errors from testing the Generalized Complex Exponential Operator (GCEO), which was trained with different training sets, using test sets of various sizes. The red dashed line indicates the results when the number of test sets matches the number of training sets. (b) shows the computational errors for different numbers of layers, using the same training set size of 360 as in the experiments.

Figure 4(b) illustrates the results of executing Euler's formula complex exponential computations using diffractive neural networks with varying numbers of layers. Increasing the neural layers from 1 to 2 and then to 3 significantly reduces the average computation error due to the increased number of neurons. However, using more than 3 layers may not further reduce the error due to transmission losses and other factors; it may even lead to an increase in the average error.

Furthermore, the generalization of the GCEO were validated in the experiment. The trained DONN successfully computed a series of angles in experiments. It is noteworthy that compared to the 360 evenly divided angles used in the training set, the 512 uniformly divided angles we employed showed a greater deviation from the 720 evenly divided angles. Despite this, the average error remained constant in these conditions. This indicates that our computational error has been effectively constrained. Further reduction in computational error could potentially be achieved by increasing the number of layers or the quantity of neurons in each layer.

Beyond processing a single input angle, our operator network design also facilitated the addition or subtraction of Euler's formula complex exponents when multiple inputs coexisted within the system, theoretically (e.g., computing the Euler's formula of an input angle plus $\pi$ results in the negation of that angle's value).

**Table 1.** Results of different complex exponential addition calculations in experiments.

| Type | $\theta_1$ (rad) | $\theta_2$ (rad) | Theoretical value | Experimental value | Relative error |
|---|---|---|---|---|---|
| $e^{i\varphi_1} + e^{i\varphi_2}$ | 2.0857 | -1.3875 | -0.3102 + 0.1129i | -0.3165 + 0.1155i | 0.0207 |
| $e^{i\varphi_1} + e^{i\varphi_2}$ | -2.4173 | -1.6930 | -0.8708 + 1.6552i | -0.8362 + 1.6910i | 0.0267 |
| $e^{i\varphi_1} + e^{i\varphi_2}$ | -1.2479 | -2.0071 | -0.1053 + 1.8546i | -0.0787 + 1.8989i | 0.0278 |
| $e^{i\varphi_1} + e^{i\varphi_2}$ | -2.8362 | 2.7489 | -1.8776 - 0.0820i | -1.8778 - 0.1376i | 0.0296 |
| $e^{i\varphi_1} + e^{i\varphi_2}$ | -0.3142 | 0.4800 | 1.8381 - 0.1527i | 1.8912 - 0.1271i | 0.0320 |
| $e^{i\varphi_1} + e^{i\varphi_2}$ | -1.2479 | -2.0071 | -0.1053 + 1.8546i | -0.0805 + 1.8637i | 0.0142 |
| $e^{i\varphi_1} + 2e^{i\varphi_2}$ | 0.7330 | 1.3963 | 0.8300 - 1.1615i | 0.8139 - 1.1778i | 0.0160 |
| $e^{i\varphi_1} + 2e^{i\varphi_2}$ | 2.1817 | 3.1241 | -1.0735 - 0.8279i | -1.0799 - 0.8555i | 0.0209 |
| $e^{i\varphi_1} + 2e^{i\varphi_2}$ | -2.6005 | 0.4451 | -0.4059 + 0.2998i | -0.4152 + 0.2922i | 0.0238 |
| $e^{i\varphi_1} + 3e^{i\varphi_2}$ | 1.0647 | -2.5220 | 0.2134 - 0.6811i | 0.2120 - 0.6586i | 0.0315 |
| $e^{i\varphi_1} + 3e^{i\varphi_2}$ | -1.7541 | -2.2689 | -0.3965 + 1.2386i | -0.3665 + 1.2671i | 0.0318 |
| $e^{i\varphi_1} + 3e^{i\varphi_2}$ | 2.5656 | 1.9460 | -0.9608 - 0.8548i | -0.9624 - 0.8591i | 0.0035 |

In consideration of the interference caused by simultaneously inputting two beams into the system, we modulated the sum or difference of the complex amplitudes of the beams at two distinct angles. Subsequently, the results of these separate calculations were combined, and the coherence term was eliminated to yield Euler's formula complex exponential addition calculations. These exhibited errors comparable to those found in single calculations. Table 1 illustrates some of the experimental results of Euler's formula complex exponential addition calculations. Additionally, to account for unbalanced additions, we amplified the energy of one input by two to three times during summation. This approach still maintained the results within an acceptable error range.

In summary, we have developed GCEO based on DONN capable of computing complex exponentials. This method surpasses conventional neural networks used in optical computation, enabling the calculation of arbitrary Euler formula complex values with limited error. Post-training, GCEO can identify the azimuthal angle of an input Gaussian beam and display the corresponding light intensity after computation at different coded apertures. For experimental implementation, a spatial light modulator and mirrors were utilized to construct a three-layer DONN in the visible spectrum at 532 nm. Furthermore, the average systematic error remained below 0.05. Our approach has experimentally achieved effective complex number computations, utilizing spatial light. Notably, a similar scheme is feasible for implementation within optical chips, harboring significant potential. This presents the possibility for further miniaturization of devices at higher frequencies, potentially laying the groundwork for chip-scale ultra-fast computation and signal processing systems. We anticipate that our GCEO will significantly contribute to advancing optoelectronic integration capabilities.

## 5. Appendix

As shown in Section 2, we use DONN to process the complex exponential computing through Euler's formula, which is based on the propagation of light through several layers (3 layers in the main text) of the diffraction plane. The total number of diffractive layers is set to be $N_l$. The total number of training samples is $N$. To reduce the computing deviation, we use the mean squared error (MSE) as the loss function for the DONN,

$$\mathcal{L} = \frac{1}{N}\sum_{\alpha=1}^{N}\left|Y_{out}^{(\alpha)} - Y_{tar}^{(\alpha)}\right|^2 \tag{6}$$

where $Y_{tar}^{(\alpha)}$ is the target output data of the DONN for α-th sample.

In the training, the gradient of the loss function $\mathcal{L}$ with respect to $\varphi_n$ is the most important item to be calculated. According to Equation (6), the derivative of $\mathcal{L}$ with respect to $\varphi_n$ is

$$\begin{cases} \dfrac{\partial \mathcal{L}}{\partial \varphi_n} = \dfrac{1}{N}\sum_{\alpha=1}^{N} K\left(Y_{out}^{(\alpha)} - Y_{tar}^{(\alpha)}\right)^* \dfrac{\partial X_{out}}{\partial \varphi_n} + c.c. \\ \dfrac{\partial X_{out}^{(\alpha)}}{\partial \varphi_n} = DT_{N_l}DT_{N_l-1}\cdots\dfrac{\partial DT_n X_n^{(\alpha)}}{\partial \varphi_n} \end{cases} \quad (7)$$

In order to facilitate the calculation, we further simplify Equation (7) and get Equation (8), which is

$$\dfrac{\partial \mathcal{L}}{\partial \varphi_n} = \dfrac{1}{N}\sum_{\alpha=1}^{N}\left[P_n^{(\alpha)}\right]^T \dfrac{\partial T_n}{\partial \varphi_n} diag\left(X_n^{(\alpha)}\right) + c.c. \quad (8)$$

with

$$P_n^{(\alpha)} = D^T \prod_{\gamma=N}^{n+1}\left(T_\gamma D^T\right)\left[K\left(Y_{out}^{(\alpha)} - Y_{tar}^{(\alpha)}\right)^*\right] \quad (9)$$

In Equation (8), we divide $\dfrac{\partial \mathcal{L}}{\partial \varphi_n}$ into 3 parts, $P_n^{(\alpha)}$, $\dfrac{\partial T_n}{\partial \varphi_n}$ and $X_n^{(\alpha)}$. $P_n^{(\alpha)}$ represents the backward propagation part, $X_n^{(\alpha)}$ represents the frontward propagation part.

According to Equation (4), (8) and (9), we train the DONN for the target of complex exponential computing. The training set includes Gaussian beams with various spatial angles ($\theta$) as the input and Gaussian spots in four apertures with various powers as the complex number output. Here, complex exponentials using angles uniformly divided into 360 between -π and π. The total number of neural layers is $N_l$ = 3, and each layer has 300 × 300 optical neurons with the neuron size of 8 × 8 μm (matching the pixel size of SLM used in the experiment), and the wavelength $\lambda$ is in the visible spectrum of 532 nm. Moreover, the distance between neighboring layers $d$ is 7.95 cm, and the distance from the last layer to the output plane $d_1$ is 10.6 cm. Specifically, we set the learning rate at 0.05 and maintained a batch size of 90, consistent with 0.25N, and used the Adam optimizer with a momentum coefficient of (0.9, 0.999).

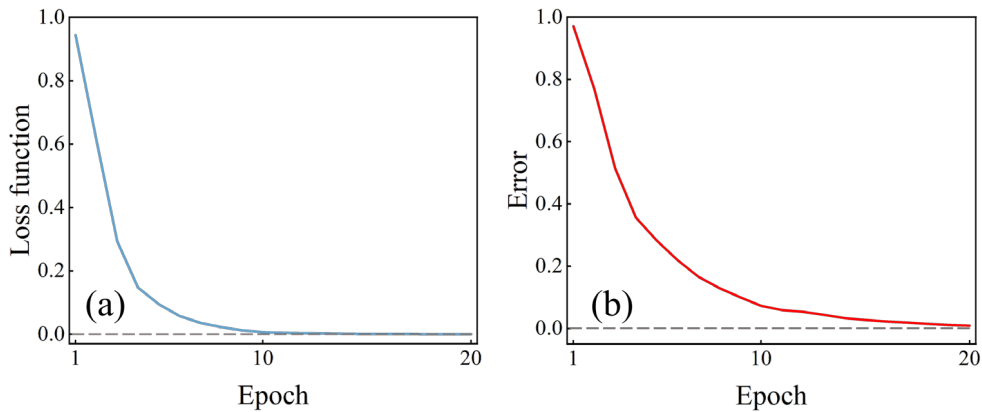

**Figure 5** Numerical results of the loss function and operating accuracy of the 3-layer DONN.

After training, the DONN can be used to compute complex exponential, and the results are shown in **Figure 5**. There, the numerical simulation of the GCEO is implemented on the platform of Python version 3.9.12 with Pytorch framework version1.11.0 (Meta. Inc.) using a desktop computer (Nvidia GeForce 2080 Super Graphical Processing Unit, GPU, and intel Core i7 12700 CPU at 2.10 GHz with 12 cores, 32 GB of RAM, running with a Microsoft Windows 10 operating system).

**Acknowledgments**

B. Chen and Q. Jia contributed equally to this work.

The authors acknowledge the funding provided by the National Natural Science Foundation of China (Grant No. 12274105), and Heilongjiang Natural Science Funds for Distinguished Young Scholars (Grant No. JQ2022A001). Fundamental Research Funds for the Central Universities (HIT.OCEF.2021020). The joint guidance project of the Natural Science Foundation of Heilongjiang Province (Grant No. LH2023A006).

**Conflict of Interest**

The authors declare no conflict of interest.

**Data Availability Statement**

The datasets generated during and/or analyzed during the current study are available from the corresponding author on reasonable request.

Received: ((will be filled in by the editorial staff))

Revised: ((will be filled in by the editorial staff))

Published online: ((will be filled in by the editorial staff))

A Generalized Complex Exponential Operator (GCEO) utilizing Diffractive Optical Neural Networks (DONN) tackles complex number computations via Euler's formula. Experimentally validated, GCEO processes over a thousand inputs with less than four percent error, demonstrating exceptional precision and generalizability. This approach breaks through data processing limitations in optical computing, promising major advances in optoelectronic integration.

Baiqiao Chen, Qi Jia,* Rui Feng, Fangkui Sun, Yongyin Cao, Jian Wang,* and Weiqiang Ding*

Generalized all-optical complex exponential operator

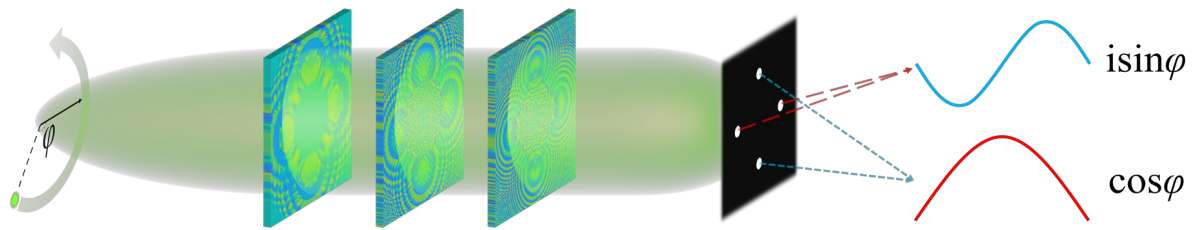